\documentstyle[aps,preprint]{revtex}
\begin{document}
\draft
\title { ENTRANCE-CHANNEL MASS-ASYMMETRY DEPENDENCE OF COMPOUND NUCLEUS
FORMATION TIME IN LIGHT HEAVY-ION REACTIONS }

\author  { A. SZANTO DE TOLEDO }

\address {\it Instituto de F\'{\i}sica da Universidade de S\~ao Paulo,
Departamento de F\'{\i}sica Nuclear-Laborat\'{\o}rio Pelletron, Caixa Postal
66318-05389-970 S\~ao Paulo, Brasil }

\author { B.V. CARLSON }

\address {\it Departamento de F\'{\i}sica, Instituto Tecnol\'ogico da
Aeron\'autica, Centro T\'ecnico Aerospacial, 12228-900 S\~ao Jos\'e dos Campos,
Brasil }

\author { C. BECK }

\address{\it Centre de Recherches Nucl\'eaires, Institut National de Physique
Nucl\'eaire et de Physique des Particules - Centre National de la Recherche
Scientifique/Universit\'e Louis Pasteur,B.P.28, F-67037 Strasbourg Cedex 2,
France }

\author { M. THOENNESSEN }

\address {\it National Superconducting Cyclotron Laboratory and Department of
Physics and Astronomy, Michigan State University,East Lansing,Michigan 48824,
USA}

\date{\today}
\maketitle

\newpage

\begin{abstract}

{\it The entrance-channel mass-asymmetry dependence of the compound nucleus
formation time in light heavy-ion reactions has been investigated within the
framework of semiclassical dissipative collision models. the model calculations
have been succesfully applied to the formation of the $^{38}$Ar compound
nucleus as populated via the $^{9}$Be+$^{29}$Si, $^{11}$B+$^{27}$Al,
$^{12}$C+$^{26}$Mg and $^{19}$F+$^{19}$F entrance channels. The shape evolution
of several other light composite systems appears to be consistent with the
so-called "Fusion Inhibition Factor" which has been experimentally observed. As
found previously in more massive systems for the fusion-evaporation process,
the entrance-channel mass-asymmetry degree of freedom appears to determine the
competition between the different mechanisms as well as the time scales
involved.}

\end{abstract}




The occurrence of an asymmetrical fusion-fission process has been identified in
light heavy-ion reactions involving dinuclear systems as light as s-d shell
nuclei \cite{Sz89} \cite{Be96}. For such light composite systems, statistically
equilibrated compound nuclei (CN) may be expected, if specific phase space
conditions are fulfilled \cite{Be95}, to coexist with dinuclear intermediate
configurations which lead to binary ``deep inelastic" collisions with strong
and fully energy damping (orbiting), and/or quasi-molecular resonant states.
Both processes may lead to very inelastic exit channels with ``isotropic"
angular distributions in the reaction plane. This coexistence has been
experimentally observed in the formation of the $^{20,22}$Ne \cite{Sz90},
$^{28}$Al \cite{An93}, $^{36}$Ar \cite{Ay88}, $^{40}$Ca \cite{Ra91}, $^{47}$V
\cite{Be93}, and $^{48}$Cr \cite{Sa95} dinuclear systems. In these studies,
most of these dinuclear systems were formed via different entrance channels, in
order to ascertain whether the CN was playing a key role in the reaction
mechanism. \\

Systematic investigations on the entrance-channel dependence and eventually on
the different time scales involved for those strongly damped processes may shed
some light on these ambiguities. However in the case of heavier systems
(A$_{CN}$ $\leq$ 160) for which noticeable time differences between the direct
component ($\tau$ $\approx$ 10$^{-21}$ sec) and compound component ($\tau$
$\approx$ 10$^{-16}$ sec) have been observed, it has been shown \cite{Th93}
that, depending on the entrance-channel mass-asymmetry parameter $\eta$ =
$\vert$(A$_{target}$-A$_{projectile}$)/(A$_{target}$+A$_{projectile}$)$\vert$ ,
the time needed for the composite system to evolve until an equilibrated CN has
been formed, may vary significantly. In this case, for long CN formation times,
the competition between the fusion process and other faster mechanisms may be
more important. As a matter of fact large time differences have been observed
in the formation of the $^{164}$Yb CN as populated by the asymmetric system
$^{16}$O + $^{148}$Sm ($\tau$ $\approx$ 10$^{-21}$ sec) or by the more
symmetric system $^{64}$Ni + $^{100}$Mo ($\tau$ $\approx$ 10$^{-20}$ sec) at
E$_{CN}^{*}$ = 49 MeV \cite{Th93}. These time differences are large enough to
induce significant divergences in the characteristic features of the CN decay.
Very recently \cite{Th95} similar time differences have been also found in the
investigations of a lighter mass region (A$_{CN}$ = 110). It was therefore
quite tempting to investigate this possibility for even lighter dinuclear
systems. \\

It is generally believed in the case of light dinuclear systems (A$_{CN}$
$\leq$ 50) that due to the smaller number of the nucleons involved (subsequent
smaller moment of inertia) and, due to the typical value of the relaxation time
\cite{Be78}, the time scales of the different binary processes involved are
strongly compressed. One of the more illustrative example can be found in the
fusion systematics which has been proposed in this mass region \cite{An90}. In
the discussion of this fusion systematics a Fusion Inhibition Factor (FIF) has
been defined and a clear correlation between this factor and the
entrance-channel mass-asymmetry parameter has been established \cite{An90}. The
FIF reflects the fact that the reaction flux which, after penetrating the
fusion barrier at a radius R$_{B}$ and a barrier height V$_{B}$ is diverted to
other reaction channels than fusion depending on the target plus projectile
combination. The FIF which is derived from the reduced fusion cross section
$\sigma$$_{red}$ = $\sigma$ E$_{cm}$ / $\pi$R$_{B}$$^{2}$, is determined from
its deviation from the linear behaviour $\sigma$$_{red}$ = (E$_{cm}$-V$_{B}$).
The tangent of the deviation angle, defined as the fraction of the flux which
penetrates the fusion barrier but does not lead to CN formation, is defined as
the Fusion Inhibition Factor. It has been clearly shown (see Fig.10 of
Ref.\cite{An90}) that FIF can be fairly well correlated with the
entrance-channel mass-asymmetry. Systems with larger initial symmetries show
larger fusion inhibition. Fig.1.a) displays the experimental FIF values (left
scale of the lower part of the figure) for light systems \cite{An90,Sz96}. It
has been suggested that this behavior might be related to the larger time scale
needed to reach the equilibrated CN configuration, allowing faster processes to
be competitive, although the linear relation is purely suggestive.\\

As soon as lighter dinuclear systems are considered, the spread in time between
the several possible processes is so much reduced that a time measurement might
not be sensitive enough to distinguish among them. Therefore time scale
measurements in the case of s-d shell composite nuclei as in the cases of the
$^{10}$B + $^{18}$O, $^{11}$B + $^{17}$O and $^{19}$F + $^{9}$Be reactions for
which the energy damped binary yields are produced dominantly by CN mechanisms
or in the case of the $^{12}$C + $^{12}$C and $^{12}$C + $^{16}$O reactions
which are capable of showing resonant processes of a quasi-molecular nature
\cite{Be95}, may not resolve the doubt whether or not the CN characteristics
keep the memory of the entrance channel properties. As a consequence, it is
still an open question whether the Bohr hypothesis is valid for very light
heavy-ion fusion reactions. Therefore, a clear understanding of the dynamics of
the collision is still lacking in this mass region. A detailed study of the
shape evolution of the dinuclear system before scission within a fusion-fission
process or involving orbiting and quasi-molecular mechanisms is highly
desirable. \\

In this paper we will propose a simple investigation of the shape evolution of
light dinuclear systems as a function of time for selected incoming angular
momenta and given entrance-channel mass-asymmetry parameters. Results are
interpreted in terms of a semiclassical dissipative collision model
\cite{Th93,Fe87}. \\

The curve of Fig.1.a) showing the experimental FIF values \cite{An90,Sz96} has
been drawn to guide the eye. Its dependence has been defined in \cite{An90} by
simple calculations of an average ``configuration lifetime" based on
proximity potentials \cite{Bl77} and on the liquid drop model \cite{CPS}. This
behaviour indicates that the entrance-channel mass-asymmetry degree of freedom
is relevant to establish a dependence of the fusion barrier and ``driving
potential" to asymetric configurations. This very qualitative result is in
quite good agreement with the conclusions reached by Thoenenssen et al.
\cite{Th93,Th95} for heavier compound systems involving equilibration times
which differ, however, to a larger extend with the initial projectile + target
combinations. \\

A more complete and sophisticated approach is based on Swiatecki's dissipative
dynamical model \cite{Sw81} which assumes full one-body dissipation. This
semiclassical model has been applied by using a particle exchange model code
HICOL written by Feldmeier \cite{Fe87}. This code \cite{Fe87} is used to follow
the evolution of the system towards equilibrium in its collective and
thermodynamic degrees of freedom. We assume that the system has reached
equilibrium in the shape degree of freedom as soon as its deformation $\beta$
does not vary significantly as a function of time. \\

The results for the four different entrance channels leading to the $^{38}$Ar
CN (at 55 MeV excitation energy) are presented in Fig.2 for L = 0, 5$\hbar$,
10$\hbar$, and 15$\hbar$ respectively. All these partial waves contribute to
the CN formation. It is shown that the most asymmetric entrance channels reach
a given stage of equilibrated deformation faster than the symmetric ones,
independently of the choice of the impact parameter (angular momentum).
Furthermore it is observed that the asymptotic value expected for the
deformation $\beta$ ( t $\rightarrow \infty$) is, in magnitude, larger with
increasing angular momentum, as expected from the Rotating Liquid Drop Model
(RLDM) (Ref.\cite{CPS}). In order to compare the results of the HICOL
calculations with that of the crude estimations of the diffusion time and FIF
values given in Fig.1.a), we have calculated the time (t) which is needed to
reach a given deformation $\beta$ (estmated to 80 percent of the final
equilibrium deformation ), starting from a given entrance-channel
mass-asymmetry (with L = 0 and L = 15$\hbar$) as shown in Fig.1b. It is
concluded that the angular momentum does not play a major role in determining
the relative drift time of the system and that in the case of a symmetric
entrance channel, the system evolves initially very fast, generating a neck and
then drifting more slowly. This might not be due essentially to the intrinsic
diffusion time but possibly to the available surface energy at this early stage
of the collision. \\

The shape evolution of the $^{38}$Ar composite system, predicted by HICOL is
depicted in Fig.3 for both the symmetric $^{19}$F + $^{19}$F and the asymmetric
$^{9}$Be+$^{29}$Si entrance channels. We observe that the equilibrium shape,
which for these systems is attained within a fraction of a revolution, can be
reached even faster for the asymmetric system. Another controversal point in
these calculations is that for frontal collisions (L = 0) nucleons drift from
the light to the heavy fragments. As soon as angular momentum is introduced
into the system, it tends to be symmetrically equilibrated. This feature can be
well understood in terms of transport phenomena as described, for instance, in
the model of the pioneer work of Randrup \cite{Ra78} which was based on the
same physical picture. More recently an alternative one-body dissipation model
including shell effects has been proposed by Bonasera \cite{Bo86}. It is
however still difficult to understand that \cite{Ta91,Ma95} from other
transport treatments of a similar approach contradictory conclusions have been
advanced in the case of more massive dinuclear systems \cite{Ma95}.
Subsequently the problem which is still open is how in the light dinuclear
systems the driving forces are overcome by the centrifugal term. \\

In conclusion, we have shown qualitatively in this Brief Report that for light
heavy-ion fusion reactions, the shape degree of freedom is expected to be
equilibrated only within a fraction of a revolution as observed in \cite{Sz90}.
As for heavier nuclear systems of the A$_{CN}$ $\approx$ 110 and 160 mass
regions \cite{Th93,Th95}, the entrance-channel mass-asymmetry degree of freedom
is found to play a key role in the time scale of the fusion process and may
determine the degree of competition between fusion and other more peripheral
binary processes such as the deep-inelastic orbiting mechanism. Experimental
attempts to measure the fission time scales involved in this light-mass region
will be soon undertaken.\\

\newpage

\centerline {\bf ACKNOWLEDGEMENTS }

\vskip 2.0 cm

This work has been partly supported by CNRS of France and CNPq of Brazil within
the framework of a CNRS/CNPq cooperation program 910106/94-0. One of us (CB)
would like to thank Y. Abe and R. M. Freeman for entlighning discussions and
for a critical reading of the manuscript.

%
%

\begin{figure}
\caption { a) (LOWER) Experimental values for the Fusion Inhibition Factor
(FIF) as defined in the text (left scale) for $^{19}$F induced reactions and
for the $^{38}$Ar CN fusion reactions. The solid curve dependence (right scale)
of the ``configuration life time" defined in Ref.13 is also given as as a
function of the mass asymmetry to guide the eye. b) (UPPER) Time scales
predicted by HICOL (see text) for the $^{19}$F+$^{19}$F, $^{12}$C+$^{26}$Mg,
$^{11}$B+$^{27}$Al and $^{9}$Be+$^{29}$Si entrance channels to reach the CN
equilibrated shape. The dashed and solid curves (drawn to guide the eyes)
represent the cases for L = 0 and L = 15$\hbar$ respectively. }
\label{FIG.1 :}
\end{figure}

\begin{figure}
\caption{ Time evolution of the nuclear deformation $\beta$, estimated to 80
percent of the final equilibrium deformation, for the $^{19}$F+$^{19}$F,
$^{12}$C+$^{26}$Mg, $^{11}$B+$^{27}$Al and $^{9}$Be+$^{29}$Si entrance channels
as predicted by HICOL, for the head-on L = 0 collision, and for the L =
5$\hbar$, L = 10$\hbar$ and L = 15$\hbar$ more peripheral collisions.}
\label {FIG.2 : }
\end{figure}

\begin{figure}
\caption { Time evolution of the nuclear shapes for the symmetric $^{19}$F+
$^{19}$F system and for the most asymmetric investigated system
$^{9}$Be+$^{29}$Si for L = 15$\hbar$. The time scale (t) is indicated in units
of 10$^{-22}$ sec.}
\label {FIG.3 :}
\end{figure}

%
%

\end{document}